\documentclass[prl,aps,twocolumn,superscriptaddress]{revtex4}
\usepackage{amsfonts}
\usepackage{dcolumn}
\usepackage{bm}
\usepackage{tikz}
\usepackage[colorlinks=true,citecolor=blue,urlcolor=blue]{hyperref}
\usepackage{amsmath,amsfonts,amssymb,times,natbib}
\usepackage[standard]{ntheorem}

\begin{document}

\title{Experimental observation of anomalous weak value without post-selection}

\author{Mu Yang}
\affiliation{CAS Key Laboratory of Quantum Information, University of Science and Technology of China, Hefei 230026, People's Republic of China}
\affiliation{CAS Center For Excellence in Quantum Information and Quantum Physics, University of Science and Technology of China, Hefei 230026, People's Republic of China}

\author{Qiang Li}
\affiliation{CAS Key Laboratory of Quantum Information, University of Science and Technology of China, Hefei 230026, People's Republic of China}
\affiliation{CAS Center For Excellence in Quantum Information and Quantum Physics, University of Science and Technology of China, Hefei 230026, People's Republic of China}

\author{Zheng-Hao Liu}
\affiliation{CAS Key Laboratory of Quantum Information, University of Science and Technology of China, Hefei 230026, People's Republic of China}
\affiliation{CAS Center For Excellence in Quantum Information and Quantum Physics, University of Science and Technology of China, Hefei 230026, People's Republic of China}

\author{Ze-Yan Hao }
\affiliation{CAS Key Laboratory of Quantum Information, University of Science and Technology of China, Hefei 230026, People's Republic of China}
\affiliation{CAS Center For Excellence in Quantum Information and Quantum Physics, University of Science and Technology of China, Hefei 230026, People's Republic of China}

\author{Chang-Liang~Ren}\email{renchangliang@cigit.ac.cn}
\affiliation{Center for Nanofabrication and System Integration, Chongqing Institute of Green and Intelligent Technology, 400714, Chinese Academy of Sciences, People¡¯s Republic of China}

\author{Jin-Shi Xu}\email{jsxu@ustc.edu.cn}
\affiliation{CAS Key Laboratory of Quantum Information, University of Science and Technology of China, Hefei 230026, People's Republic of China}
\affiliation{CAS Center For Excellence in Quantum Information and Quantum Physics, University of Science and Technology of China, Hefei 230026, People's Republic of China}
	
\author{Chuan-Feng Li}\email{cfli@ustc.edu.cn}
\affiliation{CAS Key Laboratory of Quantum Information, University of Science and Technology of China, Hefei 230026, People's Republic of China}
\affiliation{CAS Center For Excellence in Quantum Information and Quantum Physics, University of Science and Technology of China, Hefei 230026, People's Republic of China}

\author{Guang-Can Guo}
\affiliation{CAS Key Laboratory of Quantum Information, University of Science and Technology of China, Hefei 230026, People's Republic of China}
\affiliation{CAS Center For Excellence in Quantum Information and Quantum Physics, University of Science and Technology of China, Hefei 230026, People's Republic of China}

\date{\today}

\begin{abstract}
Weak measurement has been shown to play important roles in the investigation of both fundamental and practical problems. Anomalous weak values are generally believed to be observed only when post-selection is performed, i.e, only a particular subset of the data is considered. Here, we experimentally demonstrated an anomalous weak value can be obtained without discarding any data by performing a sequential weak measurement on a single-qubit system. By controlling the blazing density of the hologram on a spatial light modulator, the measurement strength can be conveniently controlled. Such an anomalous phenomenon disappears when the measurement strength becomes strong. Moreover, we find that the anomalous weak value can not be observed without post-selection when the sequential measurement is performed on each of the components of a two-qubit system, which confirms that the observed anomalous weak value is based on sequential weak measurement of two noncommutative operators. 
\end{abstract}

\maketitle

\textit{Introduction}.---
The concept of weak measurement \cite{Aharonov,Duck,Kofman,Dressel2} was introduced by Aharonov, Albert, and Vaidman in 1988. Their theory is based on the von Neumann measurement with a very weak coupling between two quantum systems. Compared to the strong measurement, weak measurement theory is a very important nonperturbative theory of quantum measurements. Over several decades of development, there have been many notable investigations from both fundamental and practical perspectives. On the one hand, weak measurement provides novel insights into a number of fundamental quantum effects, including the role of the uncertainty principle in the Hardy's paradox  \cite{Aharonov1,Lundeen1,Yokota}, the double-slit experiment \cite{Wiseman1,Mir}, the three-box paradox \cite{Resch1}, and Leggett-Garg inequalities \cite{Palacios,Dressel,Suzuki,Emary}. On the other hand, it is considered to be useful for the signal amplification while decreasing or retaining the technical noise \cite{Jordan,Viza} in parameter estimations, such as amplification measurements of small transverse \cite{Dixon,Hosten} and longitudinal shifts \cite{Brunner,Xu,Strubi}, optical nonlinearities \cite{Feizpour} and the Poynting vector field \cite{Dressel1}.

In all these applications, the strange characteristic that the weak value obtaining in the weak measurement can even exceed the eigenvalue range of a typical strong or projective measurement and is generally complex (also known as ``anomalous weak value"), is usually considered to play a vital role. The standard weak value is defined with post-selection and the anomalous weak value is usually observed by post selecting a small subset of data. Therefore the anomalous weak value is a result of post-selection, which is widely accepted in the community. However there are still great controversies on this point, especially for the validity of weak value technology in quantum metrology \cite{Ferrie,Combes,controversy}. With the in-depth research, it is shown that anomalous amplification can be realized without the requirement of post-selection of weak measurement techniques in high precise metrology protocols \cite{Rincon}. This result gives us some enlightenments, but it still does not answer the question: can anomalous weak values be attained without post-selection? Recently, a theoretical investigation pointed out that an anomalous weak value can be obtained without post-selection by sequential weak measurements \cite{Abbott}, which provided a more general insight and was deemed as ``yet another surprise"~\cite{Cohen2019}.

In this work, we experimentally obtain anomalous weak values in a sequential weak measurement without post selection, i.e., without discarding any data, in an all-optical system. The photonic polarizations and transverse momenta are chosen to be the system states and pointers, respectively. A sequential weak measurement on the product of two single-qubit observables is realized for arbitrary measurement strength controlled by the phase patterns on a liquid crystal spatial light modulator (SLM). The counter-intuitive average value of joint pointer's deflection is observed when the measurement strength is weak, while the anomalous value disappears when the measurement strength increases. Moreover, we further perform a sequential weak measurement on two observables each of which belongs to the components of a 2-qubit system and find that an anomalous weak value can never be observed based on commutative sequential weak measurements without post-selection.

\textit{Weak values without post-selection}.---The standard form of a weak value is given by,
\begin{equation}\label{weak}
\langle A_{\psi}^{\phi}\rangle_W:=\frac{\langle \phi|\hat{A}|\psi\rangle}{\langle \phi|\psi\rangle},
\end{equation}
which is the mean value of observable $\hat{A}$ when weakly measured between a preselected state $|\psi\rangle$ and
a postselected state $|\phi\rangle$ \cite{Aharonov}. As introduced in \cite{Abbott}, in particular, a trivial,
deterministic measurement of the identity operator $I$ amounts to performing no post-selection. So a weak value with no postselection, can be defined as,
\begin{equation}\label{weak-1}
\langle A_{\psi}^{I}\rangle_W:=\langle \psi|\hat{A}|\psi\rangle.
\end{equation}
Obviously the weak value with no postselection is equal to the expectation (average) value of $\hat{A}$. The result of these cases is restricted to lie in $[\eta_{\mathrm{min}}(\hat{A}), \eta_{\mathrm{max}}(\hat{A})]$ ($\eta_\mathrm{min(max)}$ represents the minimum (maximum) eigenvalue), which means anomalous weak values can not be observed. It is therefore generally considered the anomalous weak values can only be observed by post selecting a small subset of data.

However, such an interpretation is invalid for sequential weak measurements \cite{Abbott}. The sequential weak value $\langle (BA)_{\psi}^{I}\rangle_W$ with no post-selection is defined ($B$ is another observavle),
\begin{equation}\label{weak-4}
\langle (BA)_{\psi}^{I}\rangle_W:=\langle\psi|\hat{B}\hat{A}|\psi\rangle.
\end{equation}
We should note that when $\hat{A}$ and $\hat{B}$ are not commute, $\langle (BA)_{\psi}^{I}\rangle_W$ will not be contained within the interval $[\eta_{\mathrm{min}}(\hat{A},\hat{B}), \eta_{\mathrm{max}}(\hat{A}, \hat{B})]$, where $\eta_{\mathrm{min(max)}}(\hat{A},\hat{B})=\mathrm{min(max)}_{k,l}\eta_{k}(\hat{A})\eta_{l}(\hat{B})$ (${k,l}$ donate the indexes of eigenvalues). This justifies that an anomalous weak value for sequential weak measurements can occur without post-selection. (see Supplementary Materials (SM)~\cite{SM} for details)

\begin{figure}[tbp]
	\centering
	\includegraphics[width=0.9\columnwidth]{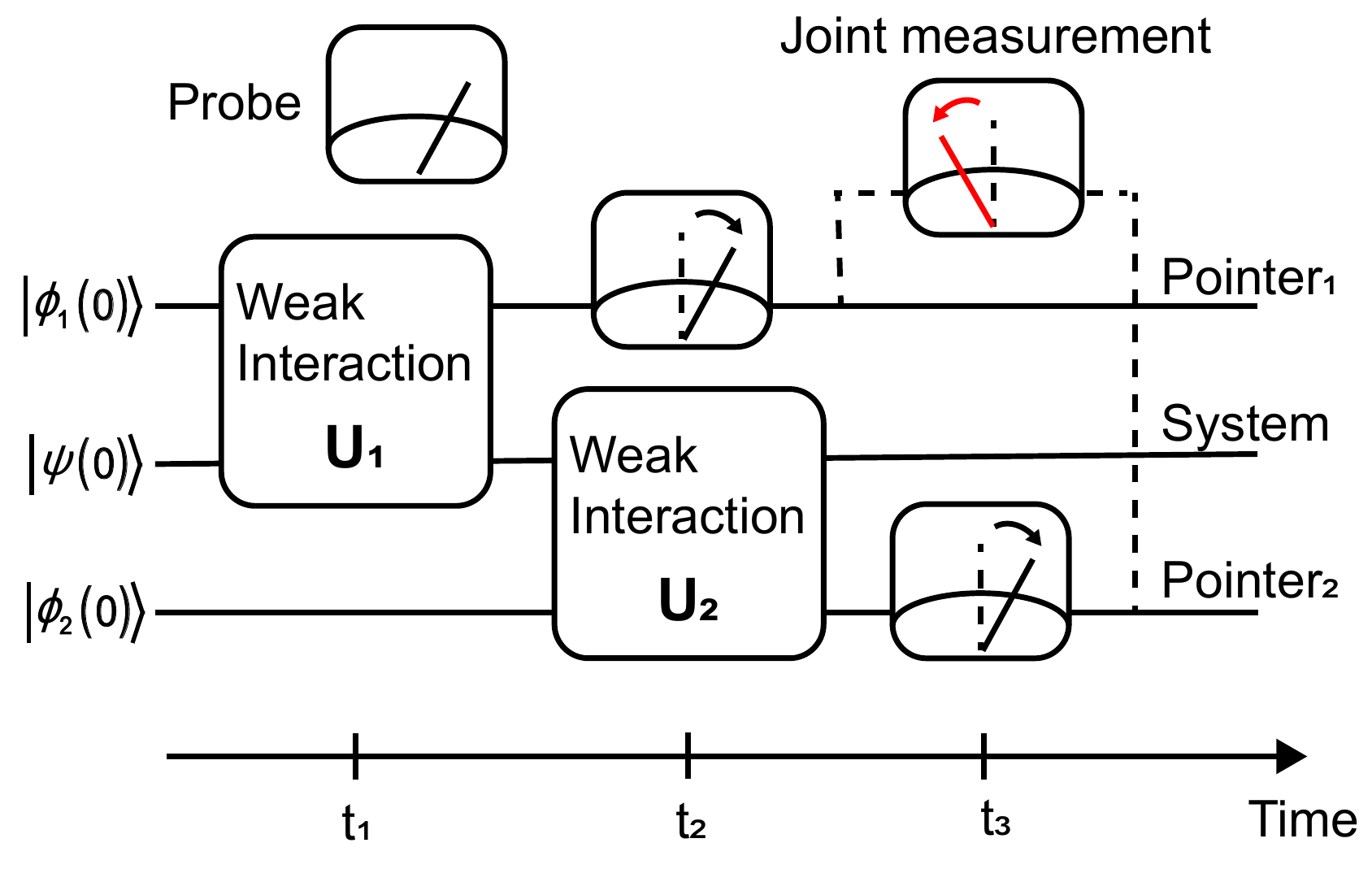}
	\caption{\label{concept}\textbf{Theoretical protocol}. The system initially in the state $|\psi(0)\rangle$ is sequencially weak coupled with two pointers in the initial states of $|\phi_1(0)\rangle$ and $|\phi_2(0)\rangle$, respectively. The time sequency is denoted as $t_1$, $t_2$ and $t_3$.  The pointers are measured individually or jointly.
	}
\end{figure}

\textit{Protocol of obtaining anomalous weak values without post-selection}.---
We consider two pointers interacting with a quantum system one after another as illustrated in Fig. \ref{concept}. Supposed a qubit system initially prepared in the general state $|\psi\rangle$, while two-pointer states prepared in $|\phi_1\rangle$ and $|\phi_2\rangle$ respectively. The evolution operator between the system and pointer are donated as $\hat{U}_i=\mathrm{Exp}[i\gamma_i\hat{A}_{i}\otimes\hat{F}_{i}]$, where $i\in\{1,2\}$ donates the $i_{th}$ weak measurement, $\gamma_i$ is the interaction coupling, $\hat{A}_{i}$ is a von Neumann measurement, and $\hat{F}_{i}$ is an operator of the pointer.  The result of the sequential weak measurements can be linked to an anomalous weak value without post-selection $\langle (A_1A_2)_{\psi}^{I}\rangle_W=\langle\psi|\hat{A}_1\hat{A}_2|\psi\rangle$.

Specifically, assume $\hat{A}_{i}$ to be the projection observable on the system with $\hat{A}_{i}=|a_i\rangle\langle a_i|$, where $|a_i\rangle=\frac{1}{2}|0\rangle-(-1)^{i}\frac{\sqrt{3}}{2}|1\rangle$, and to each measurement is associated a momentum operator $\hat{F}_{i}=\hat{p_i}$ of the pointers. Considering a Gaussian pointer with transverse wavefunction $\phi_i=\mathrm{Exp}(-x^2/\sigma_i^2)$, the joint pointer average position is given~\cite{Abbott} by
\begin{eqnarray}\label{interaction}
\langle\hat{x}_1\otimes\hat{x}_2\rangle=\mathrm{Re}[\langle (A_1A_2)_{\psi}^{I}\rangle_W]=\frac{1}{16}(1-3 e^{-\frac{\gamma_1^2}{8\sigma_1^2}})\gamma_1^2,
\end{eqnarray}
where $\hat{x}_i$ represents the position operator. Since each output pointer observable has an average shift in the range of $[0,1]$, the average value are naturally expected within the range $[0,1]$. However, when the first measurement is sufficiently weak $\gamma_1\rightarrow 0$, $\langle\hat{x}_1\otimes\hat{x}_2\rangle\approx-\frac{1}{8}\gamma_1^2$ lied out of the interval, which demonstrates the anomalous weak value. In contrast, when the first measurement is strong ($\gamma_1\rightarrow\infty$), $\langle\hat{x}_1\otimes\hat{x}_2\rangle\approx \frac{1}{16}\gamma_1^2$, which is consistent with the expectation.


\begin{figure*}[hbt!]
	\begin{center}
		\includegraphics[width=1.75 \columnwidth]{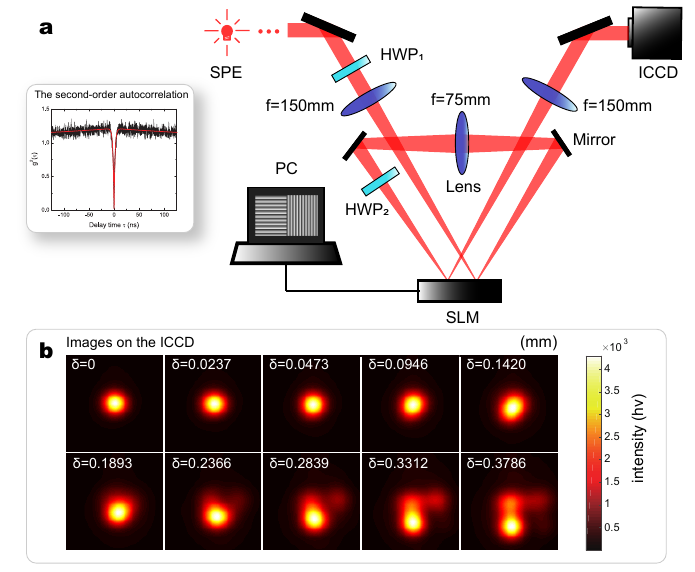}
		\caption{ \textbf{Experimental setup and deflection images.} \textbf{a}. Single photons from a single photon emitter (SPE) are sent to the sequential weak measurement setup. The single photon property is characterized by the second order autocorrelation function, in which the dip at the zero delay time is fitted to be $g^2(0)$=0.025. The polarization of the single photons are set by a half-wave plate (HWP$_1$). A lens (f=150 mm) is used to focus the photon to the right screen of the spatial light modulator (SLM) for the horizontal weak coupling, where the hologram loaded is a vertical grating. The coupling strength is adjusted by changing the density of grating. The photons are then refocused on the left screen of the SLM by a lens with f=75 mm for the vertical coupling, where the hologram loaded is a horizontal grating with the same density. The HWP$_2$ is used to rotate the polarization of the photon before the screen to adjust the direction of the pointer. The photons are then finally detected by an intensified charge coupled device (ICCD) in the focus plane of a lens with f=150 mm. \textbf{b}. The images of photon distributions detected by the ICCD with different coupling strengths $\delta$.
		}
		\label{All}
	\end{center}
\end{figure*}

\textit{Experimental setup and results}.---
We experimentally demonstrate sequential weak measurements~\cite{sequential} by using the SLM, in which a phase $\varphi$ ($\varphi=\alpha\vec{r}~mod~2\pi$, $\alpha\in\mathbb{N}$) that changes linearly along the $\vec{r}$ direction is implemented to the input photons. The evolution can be described as
\begin{eqnarray}\label{Hamiltonian}
\hat{U}_{r}= \mathrm{Exp}[-i\delta\hat{p}_{\vec{r}}\otimes|H\rangle\langle H|],
\end{eqnarray}
where $\hat{p}_{\vec{r}}= -i\frac{\partial }{\partial\vec{r}} $ (setting $\hbar=1$) represents the momentum operator, $|H\rangle$ represents the horizontal polarization of the photon and $\delta=k\alpha$ $(k\in\mathbb{R})$ represents the coupling strength which can be conveniently tuned by changing the blazing density of the hologram on the SLM (see Supplementary Materials (SM)~\cite{SM} for more details).

In the experiment, the photon's polarizations and transverse field coordinates $(\hat{x}, \hat{y})$ (or momenta $(\hat{p_x}, \hat{p_y})$) are chosen to be the system and pointers, respectively. The experimental setup is illustrated in Fig. \ref{All}a. Single photons (SPs) from an intrinsic defect in a GaN crystal are filtered by a single mode fiber~\cite{Li} (see SM~\cite{SM} for more details). The zero delay time of the second-order correlation function $g^2(0)$ is measured to be 0.262 and fitted to be 0.025, which clearly demonstrates the single photon property.

A half-wave plate (HWP$_1$) with the optical axis set to be $30^\circ$ is used to prepare the polarization state of the photon to be $|a_1\rangle=\frac{1}{2}|H\rangle+\frac{\sqrt{3}}{2}|V\rangle$, where $|V\rangle$ represents the vertical polarization. Then the photons are focused on the right screen of SLM for the first weak coupling by a convex lens with a focal length of 150 $mm$. The hologram loaded on the right area of SLM is a vertical blazed grating, and satisfies $\varphi_{right}(x,y)=\alpha\vec{x}~\mathrm{mod}~2\pi$ $(\alpha\in\mathbb{N}$). In the experiment, $\delta=0.0237\alpha$ (see SM~\cite{SM} for details). This progress can be defined as $\hat{U}_{1}$.

Similarly, through a lens with a focal length of 75~mm, photons are re-focused on the left screen of SLM for the second weak coupling, and the evolution becomes $\hat{U}_{2}$. The left hologram is a horizontal grating, which satisfies $\varphi_{left}(x,y)=\alpha\vec{y}~\mathrm{mod}~2\pi$. Another HWP$_2$ with the optical axis set to be $-30^\circ$ is placed before the screen to rotate $|H\rangle$ $(|V\rangle)$ to the state of $|a_{2}\rangle=\frac{1}{2}|H\rangle-\frac{\sqrt{3}}{2}|V\rangle$ $(|a_{2}^{\perp}\rangle=\frac{\sqrt{3}}{2}|H\rangle-\frac{1}{2}|V\rangle)$. The third Fourier lens is used to translate the photons back to the coordinate space, which are directly detected by an intensified charge coupled device (ICCD) without post-selection of  polarizations.


\begin{figure}[t!]
	\begin{center}
		\includegraphics[width=0.9 \columnwidth]{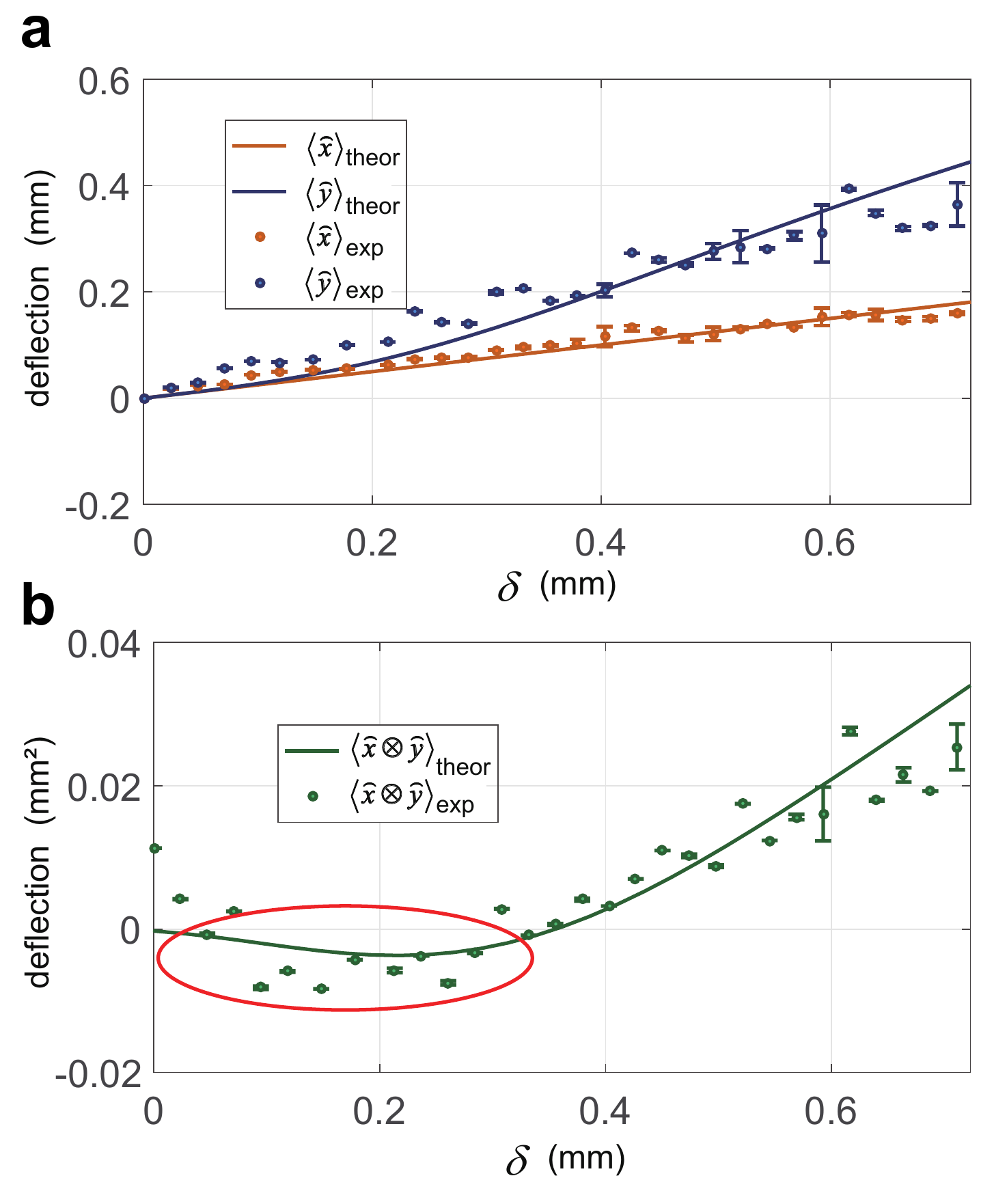}
		\caption{\textbf{The deflections of the pointer's position after sequential weak measurements in the 1-qubit system.} \textbf{a.} The brown and blue dots represent the experimental results of the pointer positions $\langle\hat{x}\rangle$ and $\langle\hat{y}\rangle$, respectively. The brown and blue lines represent the corresponding theoretical predictions.		
			\textbf{b.}	
			The green dots represent the experimental results of the joint pointer position $\langle\hat{x}\otimes\hat{y}\rangle$ with the green solid line representing the corresponding theoretical prediction. Data in the red circle represent the anomalous values.
		}
		\label{pro}
	\end{center}
\end{figure}


\begin{figure}[t!]
	\begin{center}
		\includegraphics[width=0.935 \columnwidth]{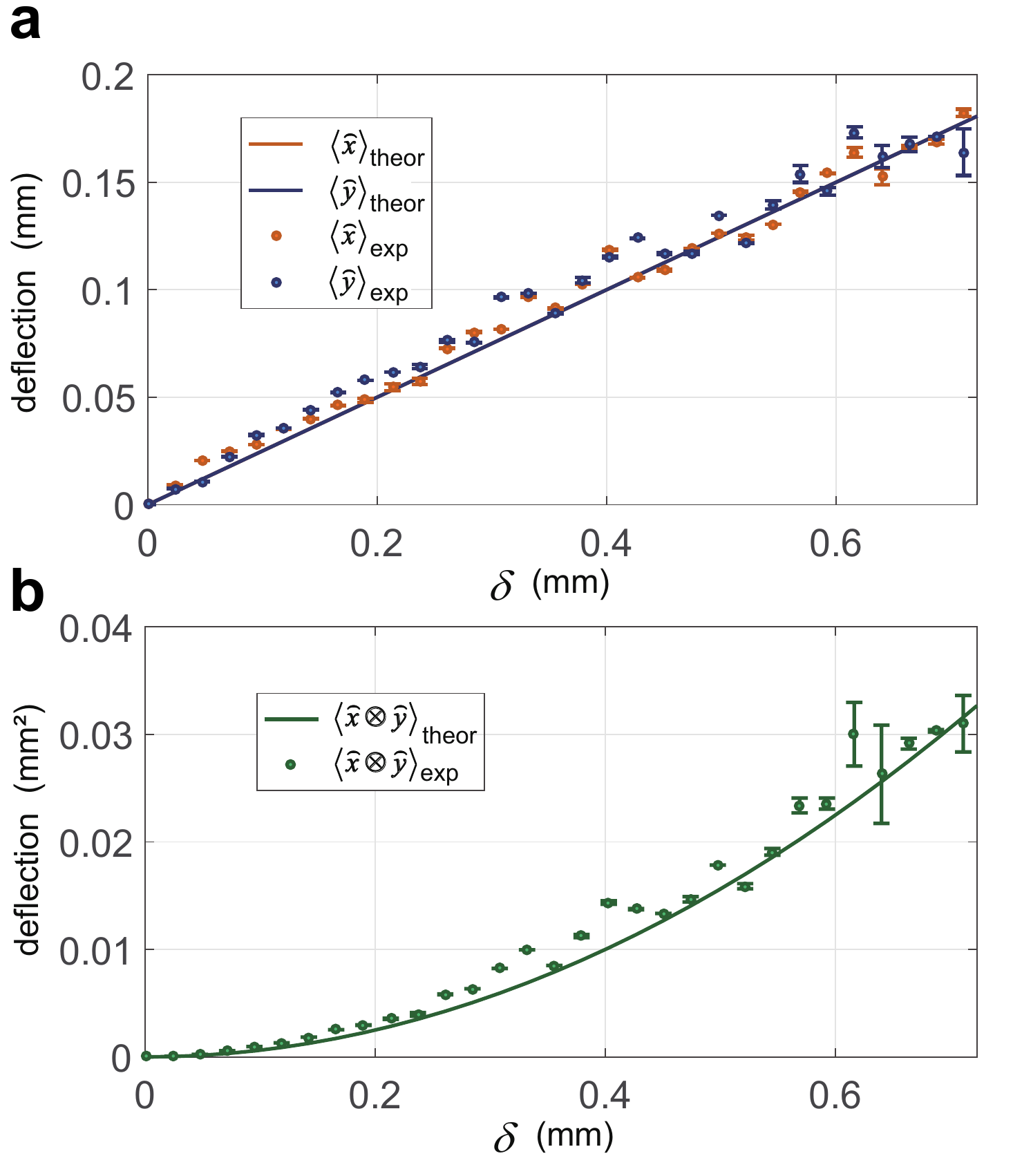}
		\caption{\textbf{The deflections of pointer positions via sequential weak measurements in the 2-qubit system.} \textbf{a.} The brown and blue dots represent the experimental results with the brown and blue lines representing the corresponding theoretical predictions, respectively.
		\textbf{b.} The green dots represent the joint average pointer positions $\langle\hat{x}\otimes\hat{y}\rangle$ with the green line representing the corresponding theoretical prediction.			
		}
		\label{non}
	\end{center}
\end{figure}

Fig. \ref{All}b shows several spatial distributions of the photons with different coupling strengths.  The scale of images is $1024\times1024$ and the pixel length is $13.5~\mu m$. The coupling strength $\delta$ varies from $0\to0.37~mm$.
When $\delta$ is less than $0.1~ mm$, the coupling strength is so small that the transverse coordinate deflection is much less than the Rayleigh distance. With the increase of coupling strength, the deflections of the transverse coordinates are clearly observed, including downward deflection $\phi^{'}(x+\delta, y)$, rightward deflection $\phi^{'}(x, y+\delta)$ and down-rightward deflection $\phi^{'}(x+\delta, y+\delta)$ after sequential weak measurements $\hat{U}_{1}$ and $\hat{U}_{2}$, respectively (see SM~\cite{SM} for more details).

To quantitatively describe the anomalous weak values, we define the average weak value $\langle \hat{R}_{\psi}^{I}\rangle_W$ (transverse coordinate pointer's positions $\langle\hat{K}\rangle$) as the mean of the transverse field coordinates,
\begin{eqnarray}\label{Mean}
\langle \hat{R}_{\psi}^{I}\rangle_W=\langle\hat{K}\rangle=\dfrac{\sum\limits_{x,y} K\cdot|\psi(x,y)|^{2} }{\sum\limits_{x,y} |\psi(x,y)|^{2}},
\end{eqnarray}
where $\hat{R}(\hat{K})\in\{\hat{A_1} (\hat{x}), \hat{A_2} (\hat{y}), \hat{A_1}\hat{A_2} (\hat{x}\otimes\hat{y})\}$. $\psi(x,y)$ is the photonic transverse wave function and $\hat{x}\otimes\hat{y}$ represents the joint pointer position operator.

Fig. \ref{pro} shows the deflection of the pointer as a function of the coupling strength. The coupling strength $\delta$ varies from $0$ to 0.711 $mm$. The pointer positions of horizontal and vertical transverse coordinates ($\langle\hat{x}\rangle$ and $\langle\hat{y}\rangle$) are shown in fig. \ref{pro}a with the brown and blue dots representing the experimental results and the brown and blue lines representing the corresponding theoretical predictions, respectively. No matter what the coupling strength is, we can find that  $\langle\hat{x}\rangle$ and  $\langle\hat{y}\rangle$ is larger than 0.
The joint pointer positions $\langle\hat{x}\otimes\hat{y}\rangle$ are shown in fig. \ref{pro}b. Anomalous phenomenon $(\langle\hat{x}\otimes\hat{y}\rangle<0)$ can be observed when the coupling strength $\delta$ is less than $0.331~mm$, which is shown in the red circle. The green dots represent the experimental results with the green line representing the corresponding theoretical prediction. When $\delta=0.189~mm$, the reversed deflection of the joint pointer reaches maximum. The experimental anomalous weak values deviates from the theoretical prediction a bit, which may due to the optical aberration of the Gaussian beams. The error bars represent the corresponding standard deviation.

Moreover, as a comparison, considered the situation of a sequential weak measurement of two observables, which are measured on each of two qubits respectively, there is no such anomalous deflection at all. Supposed a sequential measurement of the commuting observables $\hat{A}_1\otimes I$ and $I\otimes\hat{A}_2$ are performed on a bipartite system $|\psi_{12}\rangle$, the average weak value is equal to the expectation value of $\hat{A}_1\otimes\hat{A}_2$ in the absence of any post-selection, with $\langle(\hat{A}_1\otimes\hat{A}_2)_{\psi_{12}}^I\rangle_{W}=\langle\psi_{12}|\hat{A}_1\otimes\hat{A}_2|\psi_{12}\rangle$, which can never exceed the range of eigenvalues of the observable (see SM~\cite{SM} for more details).%

We experimentally investigate the case of weak measurement on two individual photons, which corresponds to the sequential weak measurement on different parts of a bipartite system.
We first apply the $\hat{U}_{x}$ to a photon and obtain the deflection of the pointer's positions in the $x$ direction ($\langle\hat{x}\rangle$). We then implement the $\hat{U}_{y}$ to the other photon and obtain the deflection of the pointer's positions in the $y$ direction ($\langle\hat{y}\rangle$). The experimental results of $\langle\hat{x}\rangle$ (brown dots) and $\langle\hat{y}\rangle$ (blue dots) with the corresponding theoretical predictions representing as brown and blue lines, respectively, are shown in fig. \ref{non}a.
The experimental results of the joint pointer deflection ($\langle\hat{x}\otimes\hat{y}\rangle$) (green dots) and the corresponding theoretical prediction (green line) are shown in fig. \ref{non}b, in which the values of $\langle\hat{x}\otimes\hat{y}\rangle$ are all larger than 0.
Error bars represent the corresponding standard deviations. All position deflections are lager than zero and satisfy $\langle\hat{x}_{A_1}\rangle=\langle\hat{y}_{A_2}\rangle=\sqrt{\langle\hat{x}_{A_1}\otimes\hat{y}_{A_2}\rangle}$. There is not any anomalous phenomenon for sequential weak measurement with any coupling strength for two qubit systems, since the two observables are commutative in the sequential case. It confirms that anomalous negative deflection observed above is based on sequential weak measurement of two noncommutative operators.

\textit{Conclusion}.---
 We have experimentally carried out a sequential measurement of two observables in a single-qubit system and a 2-qubit system for arbitrary measurement strength with the SLM. The anomalous weak values are obtained without post-selection, in which the paradox of average pointer deflection is successfully observed in a single-qubit system.
 The experimental data are coincident with the theoretical predictions. Such anomalous phenomenon disappears when the measurement strength becomes strong. On the other hand, when a sequential weak measurement is commutative, the anomalous weak value can never be observed without post-selection.

Our work clarifies a controversial point concerning abnormal weak values and post-selection.
Recently, weak values and sequential weak measurements have played crucial roles in understanding fundamental problems such as the information paradox in black holes~\cite{Horowitz, Y. Aharonov, E. Cohen, R. Bousso}, time~\cite{Aharonov2} and geometric phase~\cite{Cho}. The sequential measurement technology demonstrated in this work may be useful for exploring a series of fundamental physical concepts.

\begin{acknowledgments}
\textit{Acknowledgement}.---
We acknowledge insightful discussions with A. A. Abbott and N. Brunner.
This work was supported by the National Key Research and Development Program of China (Grant No. 2016YFA0302700), the National Natural Science Foundation of China (Grants No. 61725504, 61327901, 61490711, 11774335 and 11821404), the Key Research Program of Frontier Sciences, Chinese Academy of Sciences (CAS) (Grant No. QYZDY-SSW-SLH003), Science Foundation of the CAS (No. ZDRW-XH-2019-1), Anhui Initiative in Quantum Information Technologies (AHY060300 and AHY020100), the Fundamental Research Funds for the Central Universities (Grant No. WK2030380017 and WK2470000026).
C. L. R. acknowledge the support from National key research and development program (No. 2017YFA0305200), the Youth Innovation Promotion Association (CAS) (No. 2015317), the National Natural Science Foundation of China (No. 11605205), the Natural Science Foundation of Chongqing (No. cstc2015jcyjA00021, cstc2018jcyjAX0656), the Entrepreneurship and Innovation Support Program for Chongqing Overseas Returnees (No.cx2017134, No.cx2018040), the fund of CAS Key Laboratory of Microscale Magnetic Resonance, and the fund of CAS Key Laboratory of Quantum Information.\\
\end{acknowledgments}


\end{document}


\renewcommand{\thefigure}{S\arabic{figure}}
\renewcommand\theequation{S\arabic{equation}}

\title{Supplemental Material of `` Experimental observation of anomalous weak value without post-selection"}	

\author{Mu Yang}
\affiliation{CAS Key Laboratory of Quantum Information, University of Science and Technology of China, Hefei 230026, People's Republic of China}
\affiliation{CAS Center For Excellence in Quantum Information and Quantum Physics, University of Science and Technology of China, Hefei 230026, People's Republic of China}

\author{Qiang Li}
\affiliation{CAS Key Laboratory of Quantum Information, University of Science and Technology of China, Hefei 230026, People's Republic of China}
\affiliation{CAS Center For Excellence in Quantum Information and Quantum Physics, University of Science and Technology of China, Hefei 230026, People's Republic of China}

\author{Zheng-Hao Liu}
\affiliation{CAS Key Laboratory of Quantum Information, University of Science and Technology of China, Hefei 230026, People's Republic of China}
\affiliation{CAS Center For Excellence in Quantum Information and Quantum Physics, University of Science and Technology of China, Hefei 230026, People's Republic of China}

\author{Ze-Yan Hao }
\affiliation{CAS Key Laboratory of Quantum Information, University of Science and Technology of China, Hefei 230026, People's Republic of China}
\affiliation{CAS Center For Excellence in Quantum Information and Quantum Physics, University of Science and Technology of China, Hefei 230026, People's Republic of China}

\author{Chang-Liang Ren}\email{renchangliang@cigit.ac.cn}
\affiliation{Center for Nanofabrication and System Integration, Chongqing Institute of Green and Intelligent Technology, 400714, Chinese Academy of Sciences, People¡¯s Republic of China}

\author{Jin-Shi Xu}\email{jsxu@ustc.edu.cn}
\affiliation{CAS Key Laboratory of Quantum Information, University of Science and Technology of China, Hefei 230026, People's Republic of China}
\affiliation{CAS Center For Excellence in Quantum Information and Quantum Physics, University of Science and Technology of China, Hefei 230026, People's Republic of China}

\author{Chuan-Feng Li}\email{cfli@ustc.edu.cn}
\affiliation{CAS Key Laboratory of Quantum Information, University of Science and Technology of China, Hefei 230026, People's Republic of China}
\affiliation{CAS Center For Excellence in Quantum Information and Quantum Physics, University of Science and Technology of China, Hefei 230026, People's Republic of China}

\author{Guang-Can Guo}
\affiliation{CAS Key Laboratory of Quantum Information, University of Science and Technology of China, Hefei 230026, People's Republic of China}
\affiliation{CAS Center For Excellence in Quantum Information and Quantum Physics, University of Science and Technology of China, Hefei 230026, People's Republic of China}
\maketitle	

\section{I. Analysis for a weak value with no post-selection }
{The standard form of a weak value is given by,
\begin{equation}\label{weak}
A_{\psi}^{\phi}=\frac{\langle \phi\mid \hat{A}\mid \psi\rangle}{\langle \phi\mid \psi\rangle},
\end{equation}
which should be understood as the mean value of observable $\hat{A}$ when weakly measured between a preselected state $\mid \psi\rangle$ and
a postselected state $\mid \phi\rangle$ \cite{Aharonov}. The definition (\ref{weak}) of a weak value can be generalised
to post-selections on a given result for any general quantum measurement. As introduced in \cite{Abbott}, in particular, a trivial,
deterministic measurement of the identity operator $I$ amounts to performing no post-selection. So a weak value with no postselection, can be defined as,
\begin{equation}\label{weak-1}
A_{\psi}^{I}:=\langle \psi\mid \hat{A}\mid \psi\rangle.
\end{equation}
Obviously the weak value with no postselection is equal to the expectation value of $\hat{A}$. Moreover, the expectation value of $\hat{A}$ also equal to the sum of weak values by postselected in the projective states of a complete basis,
\begin{equation}\label{weak-2}
\langle \psi\mid \hat{A}\mid \psi\rangle=\sum_{i}\mid\langle\psi\mid i\rangle\mid^{2} A_{\psi}^{i}.
\end{equation}
The results of these three cases are equal in mathematics. But we emphasise that these correspond to different physical situations which can be easily classified which also mentioned in \cite{Abbott}. In particular, without post-selection no data is discarded. But the latter two cases \cite{Diosi} both need carry out post-selecting process on the system finally. Obviously, the way of measurements and statistics of these three cases are distinct. }

The result of these three cases is restricted to lie in $[\eta_{\mathrm{min}}(\hat{A}),\eta_{\mathrm{max}}(\hat{A})]$ ($\eta$ represents the eigenvalue). The phenomenon of a weak value outside the spectrum of $\hat{A}$ is referred to as an ¡°anomalous weak value¡±. While the notion of an anomalous weak value (\ref{weak}) for single (non-sequential) weak measurements occurs only when a non-trivial post-selection is performed, i.e., only a particular subset of the data is considered. With this in mind and given the rudimentary analysis above, it is rather natural to attribute the origin of anomalous weak values to the presence of post-selection; this opinion indeed seems to be widely shared in the community.

However, such interpretation is invalid for sequential weak measurements \cite{Abbott}. In analogy to (\ref{weak}), the sequential weak value $(BA)_{\psi}^{\phi}$ is defined, for a system prepared in $\mid \psi\rangle$ and a postselected in $\mid \phi\rangle$, as,
\begin{equation}\label{weak-3}
(BA)_{\psi}^{\phi}=\frac{\langle \phi\mid \hat{B}\hat{A}\mid \psi\rangle}{\langle \phi\mid \psi\rangle}.
\end{equation}
The sequential weak value with no post-selection can be given as
\begin{equation}\label{weak-4}
(BA)_{\psi}^{I}:=\langle \psi\mid \hat{B}\hat{A}\mid \psi\rangle.
\end{equation}
Crucially, although for a single measurement without postselection $A_{\psi}^{I}$ is simply the expectation value of $\hat{A}$, no such interpretation can be given to $(BA)_{\psi}^{I}$. When $\hat{A}$ and $\hat{B}$ are not commute, $(BA)_{\psi}^{I}$ will not be contained within the interval $[\eta_{\mathrm{min}}(\hat{A},\hat{B}),\eta_{\mathrm{max}}(\hat{A},\hat{B})]$, where $\eta_{\mathrm{min(max)}}(\hat{A},\hat{B})=\mathrm{min(max)}_{k,l}\eta_{k}(\hat{A})\eta_{l}(\hat{B})$. This justifies that an anomalous weak value for sequential weak measurements can occur without post-selection.


\section{II. Weak measurement based on liquid crystal spatial light modulator }

\begin{figure*}[hb!]
	\begin{center}
		\includegraphics[width=1 \columnwidth]{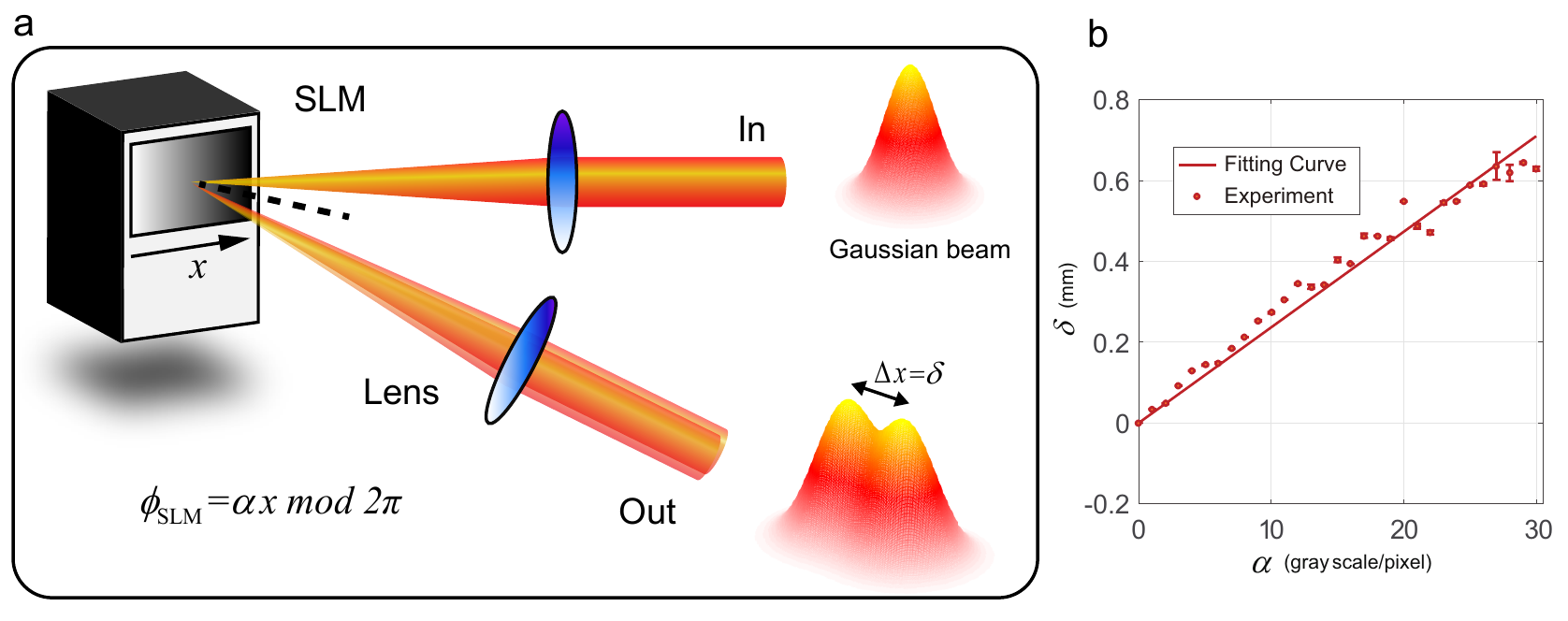}
		\caption{Weak measurement based on the liquid crystal spatial light modulator (SLM). \textbf{a.} The input photons are transformed from the coordinate space to the momentum space by a Fourier lens and focused on the screen of SLM.  A phase that changes linearly along the x direction applied on photons by the SLM, and then photons are re-transformed from the momentum space to the coordinate space by another Fourier lens. The photon wave packet will be transversely shifted slightly, which is known as weak measurement. \textbf{b.} The relation between grating densities $\alpha$ on SLM and coupling strength $\delta$.
		}
		\label{SLM}
	\end{center}
\end{figure*}

FIG. \ref{SLM}a shows the process of weak measurement. The polarization of photons are treated as the system state and momenta $(\hat{p}_x, \hat{p}_y)$ (coordinates $(\hat{x}, \hat{y})$)  of the transverse field as the pointers. The initial state is prepared as
\begin{eqnarray}\label{Initial}
|\Psi\rangle= |\varphi(x, y)\rangle(a|H\rangle+b|V\rangle),
\end{eqnarray}
where $|\varphi(x, y)\rangle$ represents the photon's spatial wave function, $|H(V)\rangle$ means the photonic horizontal (vertical) polarization, ($a$, $b$) are dimensionless complex coefficients. Through the first lens, the input photons are transformed from the coordinate space into the momentum space, which is donated as
\begin{eqnarray}
\mathscr{F}[|\varphi(x, y)\rangle](a|H\rangle+b|V\rangle)\\
= |U(\eta, \xi)\rangle(a|H\rangle+b|V\rangle),
\end{eqnarray}
where $|U(\eta, \xi)\rangle$ means the wave function in momentum representation. By using the liquid crystal spatial light modulator (SLM) to add phases $\phi_{SLM}= \alpha x~mod~2\pi$ $(\alpha\in\mathbb{N})$ on the horizontal polarized photons, where grating density is determined by $\alpha$, and the state becomes
\begin{eqnarray}
 a|U(\eta, \xi)\rangle e^{i\delta\eta}|H\rangle+b|U(\eta, \xi)\rangle|V\rangle),
\end{eqnarray}
where $\delta=k\alpha$ $(k\in\mathbb{R})$ represents coupling strength. FIG. \ref{SLM}b shows the relationship between the grating parameter $\alpha$ and the coupling strength $\delta$. The coefficient $k$ can be deduced from the fitting line of $(\alpha, \delta)$, which reads as $k=0.0237$.
Though another Fourier lens, photons are re-transformed to the coordinate space, which can be expressed as
 \begin{eqnarray}
 a\mathscr{F}^{-1}[|U(\eta, \xi)\rangle e^{i\delta\eta}]|H\rangle+b|\mathscr{F}^{-1}[U(\eta, \xi)\rangle]|V\rangle\\
 = a|\varphi(x-\delta, y)\rangle|H\rangle +b|\varphi(x, y)\rangle|V\rangle.
 \end{eqnarray}
In summery, this process simulates a weak interaction evolution $\hat{U}$, which satisfies
\begin{eqnarray}\label{Hamiltonian}
\hat{U}|\varphi(x, y)\rangle(a|H\rangle+b|V\rangle)= a|\varphi(x-\delta, y)\rangle|H\rangle +b|\varphi(x, y)\rangle|V\rangle,
\end{eqnarray}
where $\hat{U}= \mathrm{Exp}[-i\delta\hat{p}_{\vec{x}}\otimes|H\rangle\langle H|]$, and $\hat{p}_{\vec{x}}= -i\frac{\partial }{\partial\vec{x}} $ (assuming $\hbar=1$) represents horizontal momentum operator.

\section{III. Mathematical description of the experimental process}

\begin{figure*}[hb!]
	\begin{center}
		\includegraphics[width=0.8 \columnwidth]{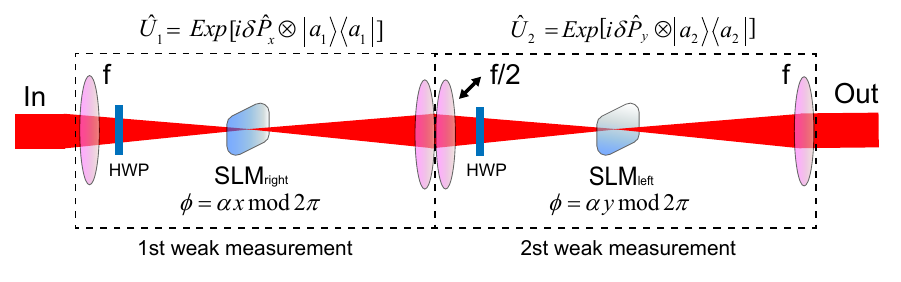}
		\caption{Concept of sequential weak measurements,  witch is equivalent to the experimental device in the main text.
		}
		\label{setup_SM}
	\end{center}
\end{figure*}

The experimental setup consists of two 4f systems, and the focal planes are located on the screen of SLM, as shown in FIG. \ref{setup_SM}. The initial photonic state are donated as
\begin{eqnarray}
|\Psi_1\rangle= |\varphi(x, y)\rangle|H\rangle,
\end{eqnarray}
where $|\varphi(x, y)\rangle=\mathrm{Exp}(-\frac{x^2+y^2}{\sigma^2})$ represents the Gaussian spatial wave function. After through a half-wave plate (HWP) with the optical axis set at $30^{\circ}$, the state becomes
\begin{eqnarray}
|\Psi_2\rangle= |\varphi(x, y)\rangle|a_1\rangle,
\end{eqnarray}
where $|a_1\rangle=\frac{1}{2}|H\rangle+\frac{\sqrt{3}}{2}|V\rangle$ $(i={1,2})$, then through the first weak measurement, the state changes into
\begin{eqnarray}
|\Psi_3\rangle= Exp(-i\delta\hat{P}_x\otimes|H\rangle\langle H|)|\Psi_2\rangle=\frac{1}{2}|\varphi(x-\delta, y)\rangle|H\rangle+\frac{\sqrt{3}}{2}|\varphi(x, y)\rangle|V\rangle),
\end{eqnarray}
Similarly, after pass the second HWP with the optical axis set at $-30^{\circ}$, we can get the state
\begin{eqnarray}
|\Psi_4\rangle= \frac{1}{2}|\varphi(x-\delta, y)\rangle|a_2\rangle-\frac{\sqrt{3}}{2}|\varphi(x, y)\rangle|a_1\rangle),
\end{eqnarray}
then through the second weak measurement,  we can get the final state
\begin{eqnarray}
|\Psi_5\rangle&=& Exp(-i\delta\hat{P}_y\otimes|H\rangle\langle H|)|\Psi_4\rangle\\
&=&-\frac{1}{4}|\varphi(x-\delta, y-\delta)\rangle|H\rangle+\frac{\sqrt{3}}{4}|\varphi(x-\delta, y)\rangle|V\rangle+\frac{3}{4}|\varphi(x, y-\delta)\rangle|H\rangle+\frac{\sqrt{3}}{4}|\varphi(x, y)\rangle|V\rangle,
\end{eqnarray}
The positions of the pointer can be expressed as
\begin{eqnarray}
\langle\hat{x}\rangle& = &\langle \Psi_4|\hat{x}|\Psi_4\rangle=\frac{1}{4}\delta \label{result1},\\
\langle\hat{y}\rangle&=&\langle \Psi_5|\hat{y}|\Psi_5\rangle=\frac{1}{8}(-3e^{-\frac{\delta^2}{8\sigma^2}}+5)\delta\label{result2},\\
\langle\hat{x}\otimes\hat{y}\rangle&=&\langle \Psi_5|\hat{x}\otimes\hat{y}|\Psi_5\rangle= \frac{1}{16}(-3e^{-\frac{\delta^2}{8\sigma^2}}+1)\delta^2, \label{result3}
\end{eqnarray}
When the measurement is sufficiently weak ($\delta\rightarrow 0$), the joint pointer deflection get anomalous, which is known as
\begin{eqnarray}\label{interaction}
\langle\hat{x}\otimes\hat{y}\rangle\approx-\frac{1}{8}\delta^2.
\end{eqnarray}
.

\section{IV. Sequential weak measurements for a 2-qubit system}

\begin{figure*}[hb!]
	\begin{center}
		\includegraphics[width=0.6 \columnwidth]{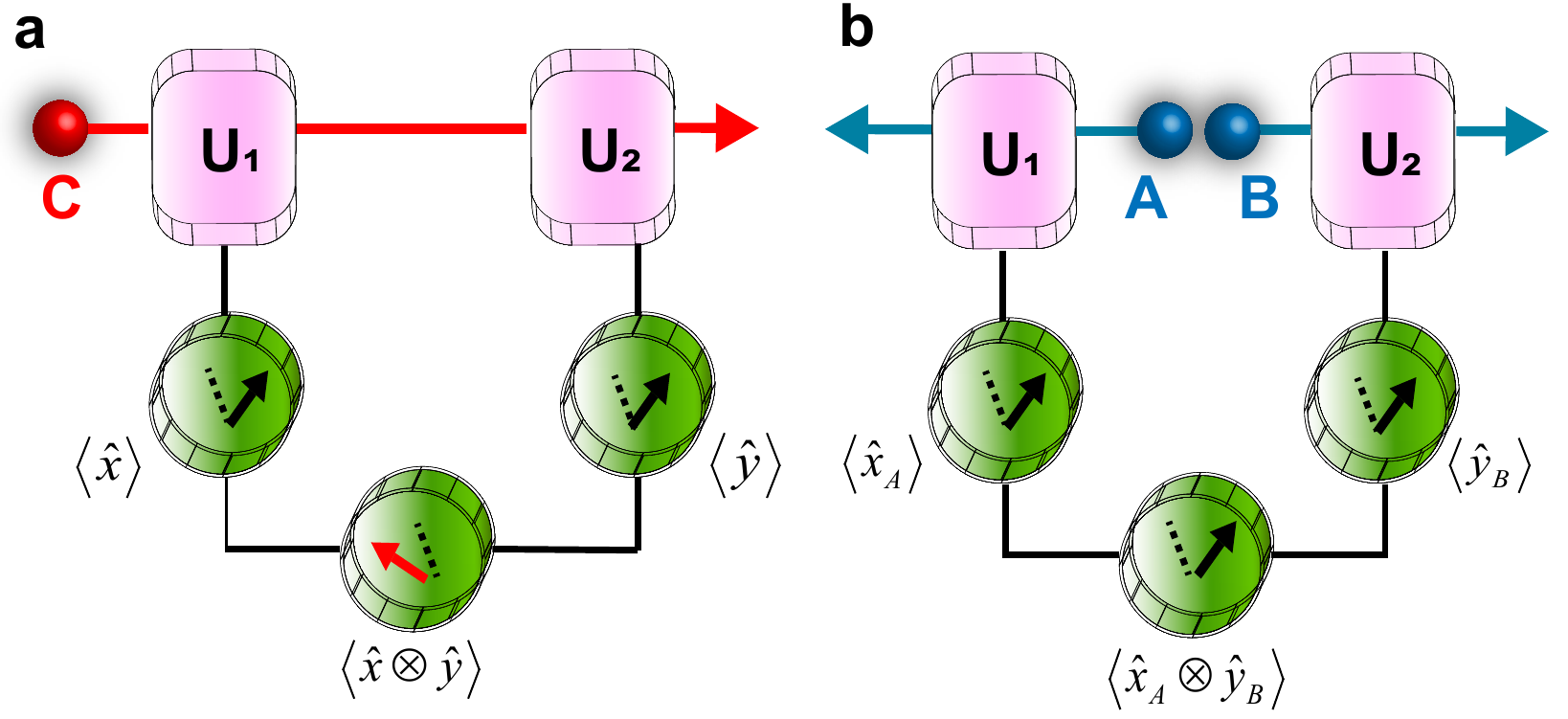}
		\caption{a. Sequential weak measurement for a single qubit system. Applying two weak measurements on the same particle C. b. Sequential weak measurement for a 2-qubit system. Two weak measurements are applied on the different particles A and B, respectively.
		}
		\label{concept}
	\end{center}
\end{figure*}


\begin{figure*}[hb!]
	\begin{center}
		\includegraphics[width=0.5 \columnwidth]{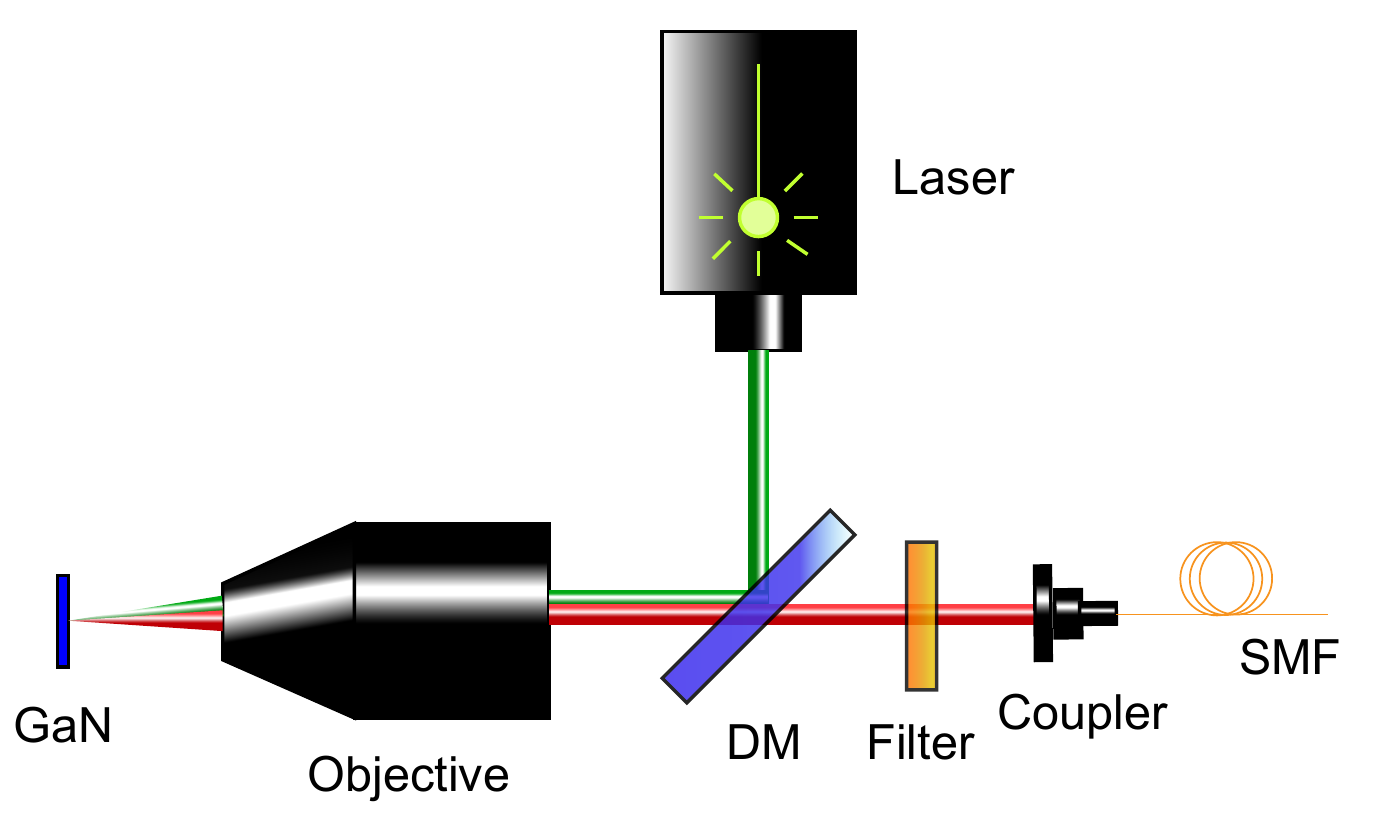}
		\caption{Experimental setup of the single photon emitter (SPE).
		}
		\label{SPE}
	\end{center}
\end{figure*}

The concept of sequential weak measurement for a single qubit system is shown in FIG. \ref{concept}a, and the pointer deflection results are shown in Eq.\ref{result1}-\ref{result3}. The concept of sequential weak measurement for a  2-qubit system is shown in FIG. \ref{concept}b. Assuming the initial states of both A and B are
\begin{eqnarray}
|\Psi_{A1}\rangle=|\varphi(x)\rangle|H\rangle\\
|\Psi_{B1}\rangle =|\varphi(y)\rangle|H\rangle
\end{eqnarray}
where $|\varphi(x(y))\rangle=\mathrm{Exp}(-\frac{x(y)^2}{\sigma^2})$. It's easy to get the final states of A and B after the weak measurement along the directions of x and y ($\hat{U}_x$ and $\hat{U}_y$), respectively, which are regarded as
\begin{eqnarray}
|\Psi_{A2}\rangle&=&-\frac{1}{4}|\varphi(x-\delta)\rangle|H\rangle+\frac{\sqrt{3}}{4}|\varphi(x-\delta)\rangle|V\rangle+\frac{3}{4}|\varphi(x)\rangle|H\rangle+\frac{\sqrt{3}}{4}|\varphi(x)\rangle|V\rangle\\
|\Psi_{B2}\rangle&=&\frac{1}{2}|\varphi(y-\delta)\rangle|H\rangle+\frac{\sqrt{3}}{2}|\varphi(y)\rangle|V\rangle
\end{eqnarray}
The positions of the pointers of A and B can be expressed as
\begin{eqnarray}
\langle\hat{x}_A\rangle& = &\langle \Psi_{A2}|\hat{x}|\Psi_{A2}\rangle=\frac{1}{4}\delta, \\
\langle\hat{y}_B\rangle&=&\langle \Psi_{B2}|\hat{y}|\Psi_{B2}\rangle=\frac{1}{4}\delta,\\
\langle\hat{x}_A\otimes\hat{y}_{B}\rangle&=&\langle \Psi_{A2}|\langle\Psi_{B2}|\hat{x}\otimes\hat{y}|\Psi_{B2}\rangle|\Psi_{A2}\rangle=\frac{1}{16}\delta^2.
\end{eqnarray}
The joint deflection of the pointer is always over than 0, so there is no anomalous phenomenon in this case.

\section{V. The preparation of the single photon emitter}

The experimental setup of the single photon emitter (SPE) is shown in FIG. \ref{SPE}.  The intrinsic defect in GaN sample is excited by a 532~nm continuous wave laser. The laser is focused by an objective with a high-NA of 0.9 after reflected by a dichroic mirror (DM). The fluorescence is collected by the same objective and is filtered by a bandpass filter with the central wave length 808~nm. The single photons are then coupled into a single mode filter (SMF), which are guided to the sequential weak measurements.

{}